# Unusual spin-wave dynamics in core-shell magnetic nanodisks

Huirong Zhao[1,2] and Ruifang Wang[1,2,3]

[1] Department of Physics, Xiamen University, Xiamen 361005, China

[2] Institute of Theoretical Physics and Astrophysics, Xiamen University, Xiamen 361005, China

[3] Collaborative Innovation Centre for Optoelectronic Semiconductors and Efficient Devices, Xiamen University, Xiamen 361005, China

Email: wangrf@xmu.edu.cn

## Abstract

We investigated the spin dynamics of a vortex state in a core-shell magnetic nanodisk driven by an oscillating field applied perpendicular to the disk plane by means of micromagnetic simulations. The nanodisk comprises a Py ($Fe_{0.2}Ni_{0.8}$) core of 100 nm in radius, surrounded by a 50 nm thick Fe shell. Fourier transform analyses show that the Py core and the Fe shell dominate spin-wave oscillation at the fundamental and higher order radial modes, respectively. For oscillating driving field tuned to the fundamental eigenfrequency, the Py/Fe interface effectively confines spin-wave excitation in the Py core region. This effect leads to significantly more rapid vortex core (VC) reversal in comparison to homogeneous disks. Our work demonstrates that the higher order modes can drive much faster VC reversal than the fundamental mode, in sharp contrast to the results obtained in homogeneous disks. With excitation levels up to 30 mT, we find strong nonlinear spin-wave dynamics in the system, which results in mode frequency redshifting, therefore the observation of the most rapid VC reversals below eigenfrequencies and VC switching in wide ranges of frequencies.

## 1. Introduction

Investigation of the spin dynamics in nanomagnets is an important subject of both fundamental physics and technological applications such as logic circuits[1-5], magnetic storage devices[1,6,7] and microwave oscillators[8-10]. In ferromagnetic nanodisks, the competition between exchange and magnetic dipole interactions favors a magnetic vortex (MV) state[3,11,12], in which the magnetization rotates in-plane around a vortex core (VC) that is magnetized perpendicular to the disk plane. This topologically nontrivial magnetic configuration is characterized by the upward or downward magnetization direction of the VC (vortex polarity $p = \pm 1$), and the counterclockwise or clockwise rotation of magnetization around the core (vortex chirality $c = \pm 1$). Experimental[6,13-17], theoretical[18-20] and numerical studies[21-24] have revealed that spin-wave excitation plays a key role in the dynamics of the polarity reversal of a MV state. These findings opened a possibility to implement vortex state nanodots in magnetic random-access memories[25,26]. A MV state is characterized by three types of spin-wave modes, namely gyrotropic[27], azimuthal[28-30] and radial[20,22,28,29]. The gyrotropic and azimuthal modes can be excited efficiently using in-plane AC magnetic field[6,14,17,31-35] or spin-polarized current[8,36,37]. In these two modes, the VC moves along a spiral path and switches its polarity through temporary creation and annihilation of a vortex-antivortex pair. The radial mode, distinct from the gyrotropical and azimuthal modes, is axially symmetric and can be excited by perpendicular AC magnetic field[22,23,38-42]. The breathing nature of radial mode forces repeated contraction and expansion of the VC. This process eventually results in a reversal of the exchange field followed by polarity switching at the core within a few picoseconds.

Recent studies[20,27-29,43] have shown that the spin-wave excitation of a MV state strongly depends on the magnetic parameters of the nanodisk. To date, the spin dynamics in homogeneous nanodisks has been intensively studied. In contrast, we have little knowledge on the dynamical behaviors of heterogeneous nanodisks, despite that heterogeneous magnetic nanostructures have demonstrated compelling spin dynamic behaviors in the studies of magnonics[44-46]. In the present work, we report markedly different spin-wave dynamics in a core-shell magnetic nanodisk under the excitation of AC magnetic field applied perpendicular to the disk plane. Our micromagnetic calculations demonstrate that higher order radial modes can drive VC reversals much faster than the fundamental mode, in sharp contrast to the case of homogeneous disks. In addition, the VC switching is observed in wide ranges of field frequency. The oscillating field creates significant nonlinear spin-wave effects that result in mode frequency redshifting, hence the most rapid VC switching events do not correspond to the mode eigenfrequencies.

## 2. Micromagnetic simulation model

The model system in study is a 20 nm thick core-shell nanodisk. The permalloy (Py: $Fe_{0.2}Ni_{0.8}$) core is 100 nm in radius. An annular iron shell is in direct contact with the Py core and is 50 nm in width (Fig. 1(a)). We use the typical Py/Fe material parameters for micromagnetic simulations[47] in this session: saturation magnetization $M_s$ = 800/1714 KA/m, exchange stiffness constant $A_{ex}$ = 13/21 pJ/m, magnetocrystalline anisotropy $K_c$ = 0/47 KJ/m$^3$ and Gilbert damping constant = 0.01/0.01. The exchange stiffness constant across the interface of Py/Fe is chosen to be $A_{ex}^{int}$ = 16 pJ/m, which is the harmonic mean of $A_{ex}$ of Fe and Py[32].

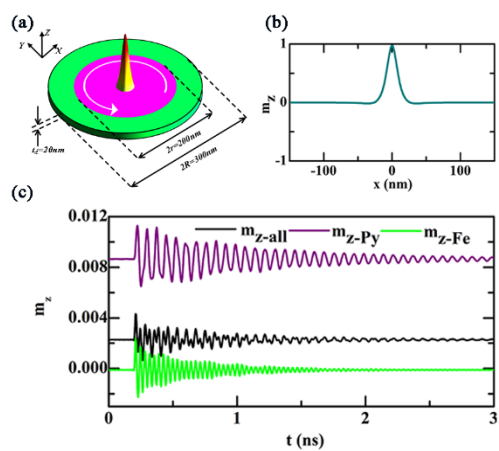

Fig. 1 (a) Schematic view of the core-shell nanodisk with indicated dimensions. The Py core (in purple) is 100 nm in radius and the *Fe* shell (in green) has a width of 50 nm. (b) The $m_z$ profile of the initial vortex state along the horizontal line across the core center. The core polarity $p$ is +1. (d) Variation of <$m_{z\text{-all}}$>, <$m_{z\text{-Py}}$> and <$m_{z\text{-Fe}}$> with time after applying a sinc-function magnetic field, which is denoted in the main text, perpendicular to the disk plane.

## 3. The radial spin-wave modes

The nanodisk is initially in a vortex state with $p = +1$, and $c = +1$ (Fig.1 (a) and (b)). The sample is then stimulated by a sinc-function field $\boldsymbol{B}(t) = \boldsymbol{e}_z B_0 \sin[2\pi f(t-t_0)]/[2\pi f(t-t_0)]$ with $B_0$ = 5 mT, $t_0$ = 0.2 ns and $f$ = 100 GHz. The oscillations of <$m_{z\text{-all}}$>, <$m_{z\text{-Py}}$> and <$m_{z\text{-Fe}}$>, which are the $z$-component magnetizations ($m_z = M_z/M_s$) averaged over the entire disk, the Py core and the Fe shell respectively, with the excitation of the sinc-function field, is shown in Figure 1(c). The much higher oscillation frequency of <$m_{z\text{-Fe}}$> than <$m_{z\text{-Py}}$> can be attributed to the significantly larger $A_{ex}$ and $M_s$ of Fe than Py. Subsequent fast Fourier transform (FFT) on <$m_{z\text{-all}}$>, <$m_{z\text{-Py}}$> and <$m_{z\text{-Fe}}$> then produces the FFT spectrums shown in Fig. 2 (a), which reveals four primary resonance peaks at $f_n$ = 13.0, 19.2, 24.0 and 27.8 GHz. The corresponding radial-spin-wave modes have indices $n$ = 1, 3, 4 and 5 respectively, which can be determined from the number of nodes along the disk radius from the FFT amplitude and phase images of the modes shown in Figure 2(b). The FFT spectrums of <$m_{z\text{-Py}}$> and <$m_{z\text{-Fe}}$> in Fig. 2(a) indicate that the Py core dominates the spin-wave oscillation at $f_1$ = 13.0 GHz, whereas the Fe shell oscillates much stronger than Py at $f_5$ = 27.8 GHz. Accordingly, the FFT amplitude images in Fig. 2 (b) illustrate strong spin-wave oscillation in the Py core ($r$ < 100 nm) but nearly zero oscillation in the Fe ring ($r$ > 100 nm) for the $n$ = 1 and $n$ = 3 modes. For $n$ = 4 and $n$ = 5 modes, we observe significant spin-wave oscillation in the Fe ring, which can be explained by the much higher $M_s$ and $A_{ex}$ of Fe than Py; thus, higher spin-wave frequency is favored in the Fe ring. The four fold symmetry in Fig 2 (b) is attributable to the cubic anisotropy of Fe. For comparison to the core-shell nanodisk, the inset of Fig. 2(a) illustrates the FFT spectrum of a homogeneous Py disk with the same overall size as the Py/Fe core-shell nanodisk. The homogeneous Py disk shows two resonance peaks at 11.3 and 15.6 GHz, which correspond to the $n$ = 1 and 3 radial modes respectively. The absence of radial modes having even indices is due to the fact that homogeneous AC magnetic field only allows for the excitation of radial modes with an odd number $n$ of nodes[23,39,41]. However, the heterogeneity of a core-shell nanodisk effectively breaks this principle, making the $n = 4$ radial mode observable.

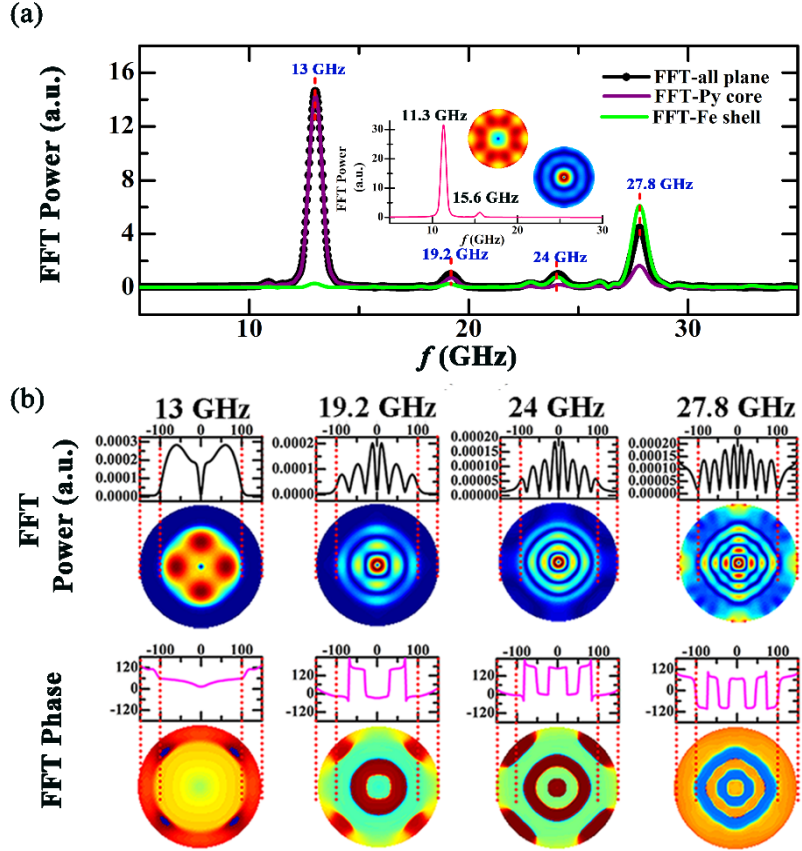

*Fig. 2.* (a) FFT Power spectrums of the damped oscillation of $<m_{z\text{-all}}>$, $<m_{z\text{-Py}}>$ and $<m_{z\text{-Fe}}>$, which was shown in figure 1(d). Four resonance peaks at frequencies of 13.0, 19.2, 24.0 and 27.8 GHz are identified, which corresponds to radial mode indices $n = 1, 3, 4, 5$ respectively. For comparison, the inset shows the FFT power spectrum, along with the FFT amplitude images, of a homogeneous Py disk with radius of 150 nm and thickness of 20 nm. (b) The FFT amplitude and phase images of the four radial spin-wave modes in the core-shell nanodisk. The images are obtained by Fourier transforming the time domain signal, $<m_{z\text{-all}}>$, recorded at each location into the frequency domain.

## 4. Dynamics of VC reversal and the nonlinear radial modes

We then applied a single harmonic magnetic field $\boldsymbol{B}(t) = -\boldsymbol{e}_z p B_0 \sin(2\pi f t)$, where $p = +1$ is the initial core polarity, $B_0 = 30$ mT is the field amplitude and $f$ is the frequency, to stimulate the core-shell disk continuously. Figure 3 shows serial snapshot images, along with the corresponding $m_z$ profiles across the core center, when $f$ equals 13.0 GHz, the $n = 1$ mode eigenfrequency. In the initial state, the VC has polarity of $+1$ and the magnetization stays in-plane elsewhere. At $t = 112$ ps, the $n = 1$ resonant field excites significant spin-wave oscillation. Then the spin-wave grows quickly in oscillation amplitude, which intensifies the periodic contraction and expansion of the VC due to the breathing nature of the radial mode. At $t = 460$ ps, the VC is compressed to a diameter of only 5 nanometers. Within the next 10 ps, the VC abruptly reverses polarity as its $m_z$ is reduced to $-1$ at $t = 470$ ps. Then high frequency spin-waves are emitted from the VC due to the release of large amount of exchange energy during the reversal period[22,38]. It is noteworthy in Fig. 3 that there is little spin-wave oscillation in the Fe ring region prior to the VC reversal, i.e., the Py/Fe interface effectively confines the spin-wave excitation in the Py core. Small amplitude spin-wave is observable in the Fe ring only after the radiation of high frequency spin-waves following the VC switching. Note that the rim of a homogeneous Py nanodisk also forms a natural boundary to confine the spin-wave[28]. However, significantly slower VC switching is reported by Dong et al.[48] for a homogeneous Py nanodisk of the same overall size as the core-shell disk in the present study. They reported VC reversal at $t = 556$ ps under the excitation of a 30 mT oscillating field that is tuned to the $n = 1$ mode eigenfrequency.

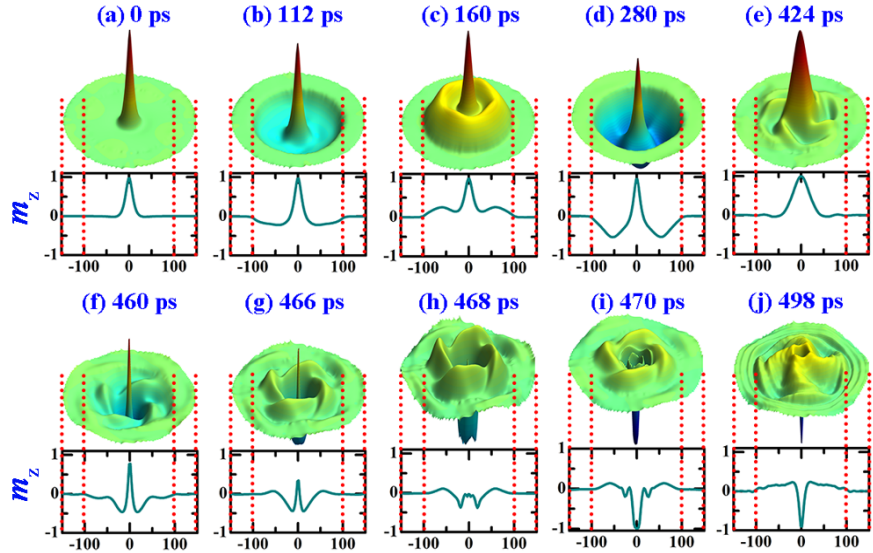

*Fig. 3.* Snapshot images showing the magnetization dynamics of the VC reversal process, after applying a perpendicular, resonant AC field tuned to the $n = 1$ mode eigenfrequency of 13 GHz. The corresponding $m_z$ profiles across the core center is presented below each snapshots.

Since the core-shell nanodisks demonstrate markedly different magnetization dynamics from the homogeneous disks, we further studied the dependence of VC reversal time ($t_{rev}$) on field frequency ($f$) in the $f$ = 10 - 30 GHz range, with variable intervals as little as 0.1 GHz (Fig. 4, blue dotted lines) . The excitation field amplitude $B_0$ is kept at 30 mT in these simulations. Figure 4(a) illustrates that the VC switching occurs in three frequency windows: $11.4 \text{ GHz} \leq f \leq 13.8 \text{ GHz}$ (W1), $18.5 \text{ GHz} \leq f \leq 18.9 \text{ GHz}$ (W2) and $22.7 \text{ GHz} \leq f \leq 27.6 \text{ GHz}$ (W3). In the U-shaped W1, the fastest VC reversals are observed in the frequency range of 11.8 to 13.1 GHz. In the rather narrow W2, $t_{rev}$ equals to 586 ps at $f$ = 18.5 GHz and quickly increases to a local maximum of 956 ps at $f$ = 18.9 GHz, which is slightly lower than the $n$ = 3 mode eigenfrequency of 19.2 GHz. Note that no VC switching is observed when the field frequency is tuned exactly to the $n$ = 3 mode eigenfrequency. In W3, $t_{rev}$ displays downward trend as $f$ increases. The $t_{rev}$ is 752 ps at $f$ = 22.7 GHz, and quickly declines to 350 ps at $f$ = 23.5 GHz before it reaches a local maximum of 426 ps at $f$ = 24.0 GHz, i.e. the $n$ = 4 resonance frequency. The $t_{rev}$ decreases to less than 300 ps as the field frequency is in the range of 24.5 GHz through 27.5 GHz. It is remarkable that the fastest VC reversal of $t_{rev}$ = 248 ps is recorded at $f$ = 27.5 GHz, which corresponds to the $n$ = 5 radial spin-wave mode. This result is in sharp contrast to previous findings obtained from homogeneous nanodisks, for which the fastest VC switching is observed only when the field frequency is tuned to the $n$ = 1 mode, and VC is increasingly more difficult to be reversed by higher modes[23,38,39].

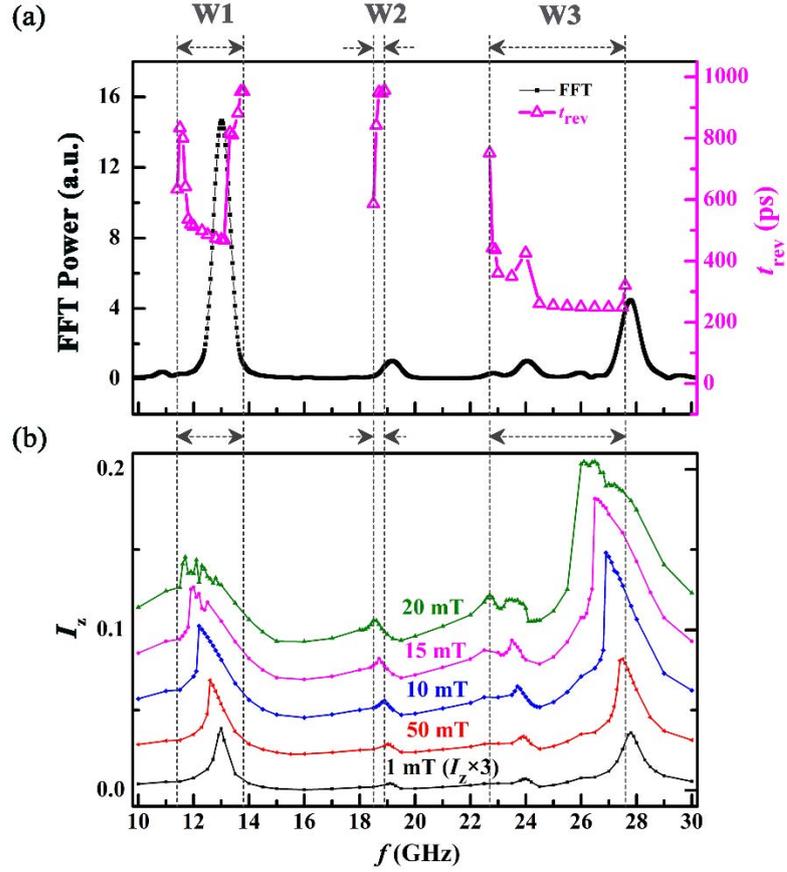

Fig. 4. (a) Variation of $t_{rev}$ with field frequency $f$. The AC external field is applied over the entire disk and the field amplitude is 30 mT. The FFT power spectrum of $<m_{z\text{-all}}>$ is also shown in order to illustrate the variation of $t_{rev}$ around the mode eigenfrequencies. (b) $I_z$, saturated oscillation amplitude of $<m_z>$, as a function of field frequency $f$, for different out-of-plane excitation amplitudes. At low field strengths, the resonance curve agrees well with a Lorentzian. When the excitation level increases, the resonance peaks show redshifting and acquire a skewness. Both phenomena are indicative of nonlinear effects.

It is noteworthy in the foregoing discussions that the local minima of $t_{rev}$ are observed below eigenfrequencies. Strangely enough, the $n$ = 3, 4 and 5 eigenfrequencies actually correspond to the local maxima of $t_{rev}$. Similar counterintuitive results were also reported in the studies of homogeneous MVs[38,39,49]. Nonetheless, the underlying mechanism of these phenomena remains unrevealed. To understand the intriguing relationship between $t_{rev}$ and $f$ around the mode eigenfrequencies, we conducted additional simulations in which sinusoidal fields are applied perpendicular to the disk plane for a duration of 8 ns. The field amplitude varies from 1 mT to 20 mT, while its frequency keeps in the range of 10 GHz to 30 GHz. The spatially averaged z-component magnetization of the sample,

<$m_z$>, oscillates under the excitation of the external field. After turning on the sinusoidal field for about 4 ns, the oscillation amplitude of <$m_z$> reaches saturation. We then recorded the saturated oscillation amplitude $I_z$, which is half the difference between the nearest maximum and minimum values of the <$m_z$> oscillation[42]. Figure 4(b) shows the variation of $I_z$ with $f$, for different excitation field amplitudes. When the external field strength is set as 1 mT, it is seen that the resonance peaks fit a Lorentzian and agree well with the FFT power spectrum displayed in Figure 2(a), since the sample is in the linear regime at the low excitation level. When the field amplitude is increased to 5 mT, the corresponding $I_z \sim f$ curve in Figure 4(b) shows that the resonance peaks shift towards lower frequencies and acquires a skewness, an indication of nonlinear behavior[39]. As the field amplitude further increases, such nonlinearity grows accordingly, i.e. the resonance peaks show stronger redshifting and leftward skewness. Such nonlinear behavior of the core-shell nanodisk is analogous to a duffing oscillator[42] attached to a nonlinear softening spring ($\beta < 0$). At the excitation strength of 20 mT, the resonance frequencies were shifted to 11.9 GHz, 18.5 GHz, 23.4 GHz, and 26.5 GHz for the $n = 1, 3, 4,$ and 5 radial modes respectively. Note that, for excitation levels of 15 mT and 20 mT, small depressions appear near the resonance frequencies, as a result of VC reversal events[39] and subsequent emission of high-frequency spin-waves. For an excitation levels above 20 mT, the $I_z \sim f$ curve is no longer obtainable because the VC reversal becomes much more rapid and the system never reaches a well-defined saturated oscillation amplitude $I_z$. We therefore conclude that the system reaches a saturation of resonance redshifting when the excitation amplitude is above 20 mT. By comparing the $t_{rev} \sim f$ plot in Fig. 4(a) with the nonlinear resonance curves obtained at excitation of 20 mT (Fig. 4(b)), one finds out that the local minima of $t_{rev}$ is in good agreement with the shifted resonance frequencies. The observation of VC reversals in wide range of frequencies in Fig. 4(a) is also attributable to the strong duffing-type nonlinear resonance of the system, under the excitation level of 30 mT.

In conclusion, we demonstrated unusual spin dynamics in a core-shell magnetic nanodisk, under the excitation of AC magnetic field applied perpendicular to the disk plane. The Py core and the Fe shell display strong spin-wave oscillation at fundamental and higher order radial modes, respectively. The markedly different VC reversal behaviors found in this heterogeneous disk can be explained by its strong nonlinear spin-wave effects.

We acknowledge the financial support from the National Natural Science Foundation of China under Grant Nos. 10974163 and 11174238.

[1]  S.-K. Kim, J. Phys. D: Appl. Phys. **43**, 264004 (2010).
[2]  D. A. Allwood, G. Xiong, C. C. Faulkner, D. Atkinson, D. Petit, and R. P. Cowburn, Science **309**, 1688 (2005).
[3]  R. Hertel, Nature Nanotech. **8**, 318 (2013).
[4]  R. Hertel, W. Wulfhekel, and J. Kirschner, Phys. Rev. Lett. **93**, 257202 (2004).
[5]  J. Lan, W. Yu, R. Wu, and J. Xiao, Phys. Rev. X **5**, 041049 (2015).
[6]  B. Van Waeyenberge *et al.*, Nature **444**, 461 (2006).
[7]  B. Pigeau, G. de Loubens, O. Klein, A. Riegler, F. Lochner, G. Schmidt, L. W. Molenkamp, V. S. Tiberkevich, and A. N. Slavin, Appl. Phys. Lett. **96**, 132506 (2010).
[8]  V. S. Pribiag, I. N. Krivorotov, G. D. Fuchs, P. M. Braganca, O. Ozatay, J. C. Sankey, D. C. Ralph, and R. A. Buhrman, Nat. Phys. **3**, 498 (2007).
[9]  K. Y. Guslienko, Journal of Spintronics and Magnetic Nanomaterials **1**, 70 (2012).
[10]  A. S. Jenkins *et al.*, Nature Nanotech. **11**, 360 (2016).
[11]  A. Wachowiak, J. Wiebe, M. Bode, O. Pietzsch, M. Morgenstern, and R. Wiesendanger, Science **298**, 577 (2002).
[12]  T. Shinjo, T. Okuno, and R. Hassdorf, Science **289**, 930 (2000).
[13]  K. Yamada, S. Kasai, Y. Nakatani, K. Kobayashi, H. Kohno, A. Thiaville, and T. Ono, Nat. Mater. **6**, 270 (2007).
[14]  B. Pigeau, G. de Loubens, O. Klein, A. Riegler, F. Lochner, G. Schmidt, and L. W. Molenkamp, Nat. Phys. **7**, 26 (2010).
[15]  T. Kamionka *et al.*, Phys. Rev. Lett. **105** (2010).
[16]  J. Park and P. Crowell, Phys. Rev. Lett. **95**, 167201 (2005).
[17]  M. Curcic *et al.*, Phys. Rev. Lett. **101**, 197204 (2008).
[18]  K. Y. Guslienko, K.-S. Lee, and S.-K. Kim, Phys. Rev. Lett. **100**, 027203 (2008).
[19]  A. Janutka and P. Gawroński, J. Phys. D: Appl. Phys. **50**, 145003 (2017).
[20]  K. Vogt *et al.*, Phys. Rev. B **84**, 174401 (2011).
[21]  M. Noske *et al.*, Phys. Rev. Lett. **117**, 037208 (2016).
[22]  R. Wang and X. Dong, Appl. Phys. Lett. **100**, 082402 (2012).
[23]  Z. Wang, M. Li, and R. Wang, NJPh **19**, 033012 (2017).
[24]  K.-S. Lee, S.-K. Kim, Y.-S. Yu, Y.-S. Choi, K. Y. Guslienko, H. Jung, and P. Fischer, Phys. Rev. Lett. **101**,

267206 (2008).

[25] Y.-S. Yu, H. Jung, K.-S. Lee, P. Fischer, and S.-K. Kim, Appl. Phys. Lett. **98**, 052507 (2011).

[26] S.-K. Kim, K.-S. Lee, Y.-S. Yu, and Y.-S. Choi, Appl. Phys. Lett. **92**, 022509 (2008).

[27] K. Y. Guslienko, B. A. Ivanov, V. Novosad, Y. Otani, H. Shima, and K. Fukamichi, J. Appl. Phys. **91**, 8037 (2002).

[28] K. Y. Guslienko, W. Scholz, R. Chantrell, and V. Novosad, Phys. Rev. B **71**, 144407 (2005).

[29] M. Buess, R. Höllinger, T. Haug, K. Perzlmaier, U. Krey, D. Pescia, M. Scheinfein, D. Weiss, and C. Back, Phys. Rev. Lett. **93**, 077207 (2004).

[30] L. Giovannini, F. Montoncello, F. Nizzoli, G. Gubbiotti, G. Carlotti, T. Okuno, T. Shinjo, and M. Grimsditch, Phys. Rev. B **70**, 172404 (2004).

[31] M. Weigand *et al.*, Phys. Rev. Lett. **102**, 077201 (2009).

[32] S. Choi, K.-S. Lee, K. Guslienko, and S.-K. Kim, Phys. Rev. Lett. **98**, 087205 (2007).

[33] M.-W. Yoo and S.-K. Kim, J. Appl. Phys. **117**, 023904 (2015).

[34] M. Sproll *et al.*, Appl. Phys. Lett. **104**, 012409 (2014).

[35] M. Kammerer *et al.*, Nat. Commun. **2**, 279 (2011).

[36] Y.-S. Choi, M.-W. Yoo, K.-S. Lee, Y.-S. Yu, H. Jung, and S.-K. Kim, Appl. Phys. Lett. **96**, 072507 (2010).

[37] W. Jin, H. He, Y. Chen, and Y. Liu, J. Appl. Phys. **105**, 013906 (2009).

[38] M.-W. Yoo, J. Lee, and S.-K. Kim, Appl. Phys. Lett. **100**, 172413 (2012).

[39] M. Helsen, A. Gangwar, J. De Clercq, A. Vansteenkiste, M. Weigand, C. Back, and B. Van Waeyenberge, Appl. Phys. Lett. **106**, 032405 (2015).

[40] M. Noske, H. Stoll, M. Fahnle, R. Hertel, and G. Schutz, Phys. Rev. B **91**, 014414 (2015).

[41] Z. Y. Wang and R. F. Wang, Aip Advances **6**, 125121 (2016).

[42] K.-W. Moon, B. S. Chun, W. Kim, Z. Q. Qiu, and C. Hwang, Scientific Reports **4**, 6170 (2014).

[43] B. Taurel, T. Valet, V. V. Naletov, N. Vukadinovic, G. de Loubens, and O. Klein, Phys. Rev. B **93**, 184427 (2016).

[44] V. Kruglyak, S. Demokritov, and D. Grundler, J. Phys. D: Appl. Phys. **43**, 264001 (2010).

[45] M. Dvornik, Y. Au, and V. Kruglyak, in *Magnonics* (Springer, 2013), pp. 101.

[46] B. Lenk, H. Ulrichs, F. Garbs, and M. Münzenberg, Phys. Rep. **507**, 107 (2011).

[47] We used LLG Micromagnetic Simulator ver. 3.14d developed by M. R. Scheinfein to carry out micromagnetic simulations.

[48] X. Dong, Z. Wang, and R. Wang, Appl. Phys. Lett. **104**, 112413 (2014).

[49] O. V. Pylypovskyi, D. D. Sheka, V. P. Kravchuk, F. G. Mertens, and Y. Gaididei, Phys. Rev. B **88**, 014432 (2013).